# Plasmonics with a twist: taming optical tornadoes on the nanoscale

**Svetlana V. Boriskina**

Mechanical Engineering Department, Massachusetts Institute of Technology, 77 Massachusetts Avenue, Cambridge, MA

**Abstract**

This chapter discusses a hydrodynamics-inspired approach to trap and manipulate light in plasmonic nanostructures, which is based on steering optical powerflow around nano-obstacles. New insights into plasmonic nanofocusing mechanisms are obtained by invoking an analogy of the 'photon fluid' (PF).  By proper nanostructure design, PF kinetic energy can be locally increased via convective acceleration and then converted into 'pressure' energy to generate localized areas of high field intensity. In particular, trapped light can be molded into optical vortices – tornado-like areas of circular motion of power flux – connected into transmission-like sequences. In the electromagnetic theory terms, this approach is based on radiationless electromagnetic interference of evanescent fields rather than on interference of propagating waves radiated by the dipoles induced in nanoparticles. The resulting ability to manipulate optical powerflow well beyond the diffraction limit helps to reduce dissipative losses, to increase the amount of energy accumulated within a nanoscale volume, and to activate magnetic response in non-magnetic nanostructures. It also forms a basis for long-range on-chip energy transfer/routing as well as for active nanoscale field modulation and switching.

## 12.1. Introduction

Noble-metal nanostructures known for their unique ability to squeeze light into sub-wavelength volumes enable a broad range of fascinating applications in optoelectronics, biomedical research, energy harvesting and conversion [1-5]. Most of these applications - Raman and fluorescence sensing being the most widespread – make use of the electric field enhancement and high local density of states provided by plasmonic nanostructures to amplify weak molecular signals such as Rayleigh [6-8] or Raman [9-13] scattering efficiency, absorption efficiency [14], fluorescence rate [15-18], etc.

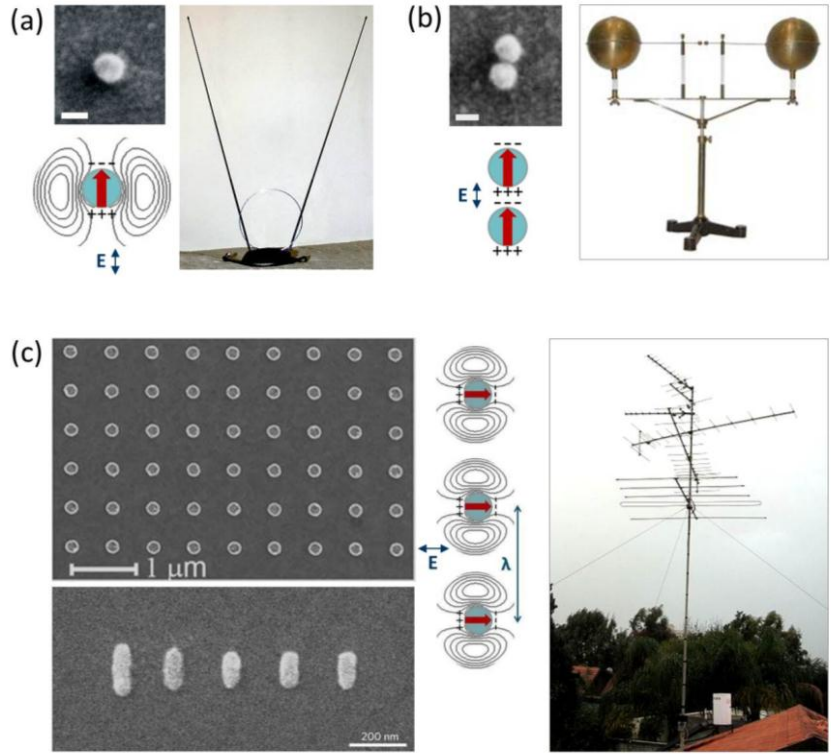

Fig. 12.1. Noble-metal nanoparticles as optical nanoantennas. (a) A single metal nanoparticle with an oscillating electric dipole moment induced by an external light source serves as an analog to an RF dipole antenna. The scale bars in SEM images here and in (b) are 50 nm (adapted with permission from [19], ©ACS). (b) Dimer-gap nanoantenna as an analog to a classical Hertzian dipole radio-frequency radiator. (c) Ordered metal nanoparticle arrays (adapted with permission from: top left [20] ©AIP, bottom left [21] ©NPG) as optical analogs to phased periodic antenna arrays and Yagi-Uda antennas. The insets show schematics of the charge distribution and emission patterns of dipole LSP modes of individual nanoparticles. Antennas images source: Wikimedia Commons.

The electric field enhancement is a manifestation of the resonant excitation of collective oscillations of free charge carriers in some materials (e.g., free electrons in metals) by the external light sources or embedded emitters. In the language of quantum mechanics, this process can be explained as the excitation of plasmons (quanta of plasma oscillations). The collective dynamics of plasmons is driven by the long-range correlations caused by Coulomb forces, and can be drastically modified by engineering the boundaries of the spatial region filled by electron plasma. In particular, bulk plasmons confined in thin metal films can couple with photons to create surface plasmon-polariton (SPP) waves [22]. On the other hand, illumination-induced collective response of plasmons confined in nanoparticles is manifested in the excitation of quantized localized surface plasmon (LSP) modes with different angular momenta [23].

Although being quantum in nature, many plasmonic effects can be well described in the frame of the classical electromagnetic theory by using a semiclassical Drude-Lorentz-Sommerfeld model to define the frequency-dependent permittivity of metals. As a result, many fundamental principles and engineering solutions established in the electromagnetic modeling of radio-frequency (RF) antennas, microwave transmission lines, circuit elements, etc. can be applied to study and design plasmonic nanostructures. These include the concepts of antenna resistance, directivity and efficiency [24] and the principle of impedance matching [25, 26] just to name a few. Overall, the striking analogies with RF and microwave engineering – which focus on reversible interfacing the propagating radiation with the fields localized inside sub-wavelength-size components – cemented the use of the design framework that treats plasmonic nanostructures as nanoscale analogs of RF antennas [17, 24, 27-29] and waveguides [2, 26, 30-32]. Alternative yet closely-related formulations are based on the treatment of plasmonic components as lumped circuit elements [33, 34] and resonators for plasmons [35-37].

The concept of an optical antenna is a very straightforward and visual analogy to describe a plasmonic nanoparticle, which can serve as both electromagnetic transmitter and receiver capable of converting the incoming radiation into localized surface plasmon modes, subsequently re-radiating it into the far-field, or modifying emission properties of embedded molecules. The oscillating dipole moments in individual particles induced by external sources can in turn be treated as secondary sources of electromagnetic radiation, i.e., dipole antennas (Fig. 12.1a). Dimer-gap nanoantenna configuration, which provides larger radiation efficiency as well as higher spectral tunability than an individual nanoparticle, can be constructed by analogy with a classical radio-frequency radiator, the Hertzian dipole (Fig. 12.1b) [27, 38]. The RF technology analogies so deeply penetrated the field of plasmonics research, that the terms 'noble-metal nanoparticle(s)' and 'optical nanoantenna(s)' became almost interchangeable. Furthermore, since the collective response of multi-particle structures is governed by the long-range interactions between individual dipoles (as long as the interparticle spacing is large enough so that the contribution of higher order resonances is negligible), various phased RF antenna array configurations have been successfully replicated on the nanoscale (Fig. 12.1c) [20, 21, 39-42].

On the other hand, for closely-packed nanoparticle clusters – where short-range interactions play the major role – a theoretical framework drawing analogies with chemistry has been successfully used. Owing to the similarities between the properties of confined quantized plasmonic states in nanoparticles and confined elec-tron states in atoms, the former can be dubbed 'plasmonic atoms'. Just as atomic wavefunctions added together produce molecular orbitals, in-phase and out-of-phase overlaps of electromagnetic fields of LSP modes in plasmonic atoms result in the splitting of their energy levels [43] (Fig. 12.2a) and hybridization into bonding and anti-bonding modes of plasmonic molecules (shown in Fig. 12.2b for the lowest-energy modes of a diatomic molecule – nanoparticle dimer). The in-phase overlap combination (a bonding mode, Fig. 12.2a, left) produces a build-up of electromagnetic energy density in the gap between the nanoparticles, while the out-of-phase overlap results in the decrease in energy density down to a zero value in the gap center for the anti-bonding mode (Fig. 12.2b, right). Also, in perfect analogy with the molecular orbitals, the anti-bonding mode is higher in energy (has a shorter wavelength), while bonding mode is lower in energy than the dipole mode of an isolated plasmonic nanoparticle (see, e.g., Fig. 12.2 in [13]). More complex nature-inspired plasmonic molecule structures – such as aromatic plasmonic molecules shown in Fig. 12.2c) have been proposed [44, 45], offering a natural extension to the concept of plasmonic crystals and quasicrystals [46, 47]. It should also be noted that this chemical analogy has been previously exploited to study coupling of confined optical states in 'photonic molecules' [48].

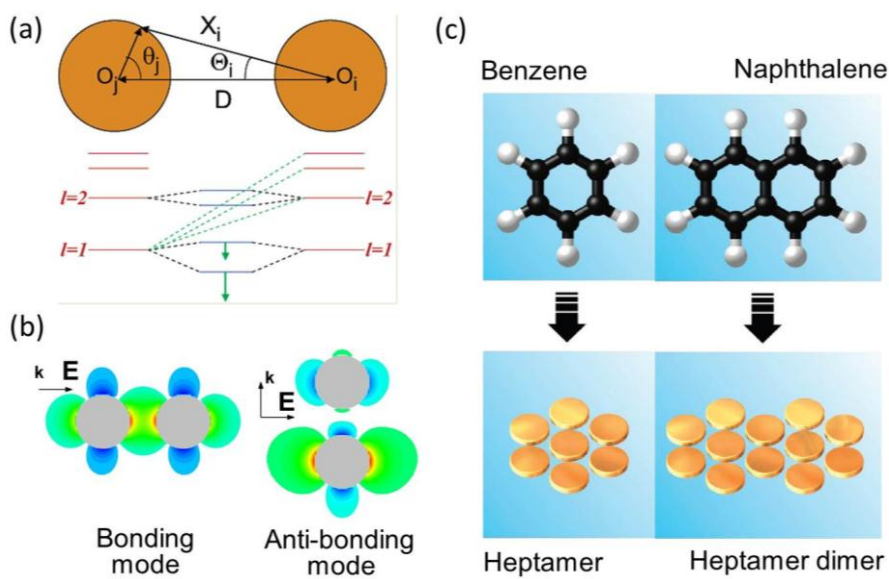

Fig. 12.2. Noble-metal nanoclusters as plasmonic molecules. (a) Splitting of the energy levels of LSP resonances in the nanoparticle dimer (reproduced with permission from [43], ©ACS). (b) Spatial distributions of the electric fields of the bonding (often referred to as longitudinal) and anti-bonding (often referred to as transverse) coupled-dipole modes. (c) Individual and coupled nanoparticle heptamers as plasmonic analogs of aromatic chemical molecules (adapted with permission from [44] ©ACS).

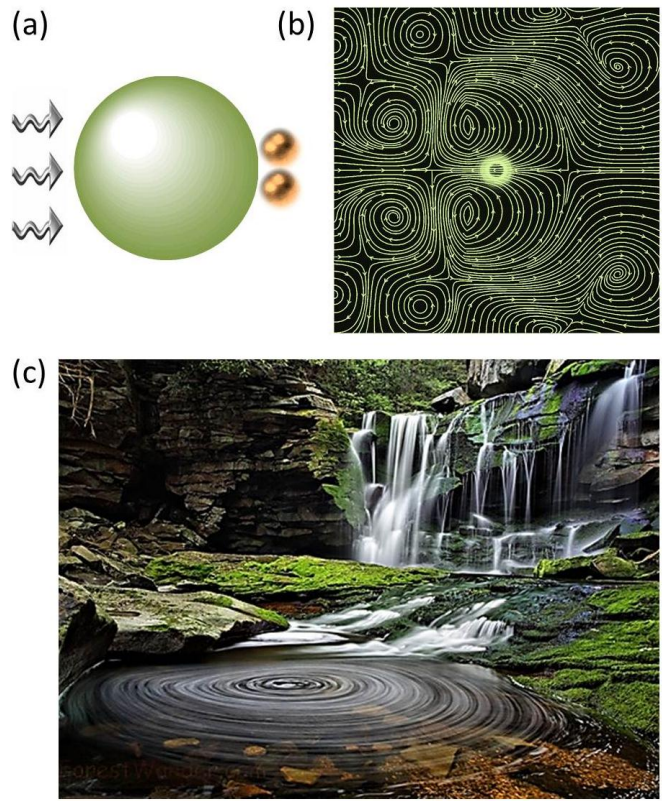

Fig. 12.3. Hydrodynamical modeling of plasmonic nanostructures: nanoparticles as the obstacles modifying the powerflow path. (a) An example of a photonic-plasmonic structure that molds optical powerflow into coupled counter-rotating vortices (shown in (b), see [49] for details). (c) An example of a complex trajectory of the fluid motion through the obstacle course, which features formation of the areas of convective flow acceleration and flow vortices [50].

In this chapter, I discuss in detail an alternative approach to the plasmonic design that does not cast the problem in the conventional terms of a sequence of scattering events but treats it as continuous energy flow process. This approach draws inspiration from hydrodynamics, which studies how the flow of fluids can be manipulated by obstacles strategically positioned in the flow path. Accordingly, the focus shifts to the studies of the Poynting vector (which is a direct analog of a fluid flux) rather than the electromagnetic field behavior under the influence of nanostructures. Many interesting hydrodynamic effects can be engineered by a proper design of the obstacles pattern, including convective acceleration or deceleration even in the case of steady flow, formation of flow vortices – areas of spinning flow of fluid – as well as flows collisions accompanied by the formation of 'shock waves.' Electromagnetic analogies of these effects can be invoked in the plasmonic design, offering new insights into routing, re-circulation, and concen-tration of optical energy within nanostructures. In particular, optical energy flow can be molded into optical vortices – tornado-like areas of circular motion of energy flux – 'pinned' to nanostructures (see Fig. 12.3). In the following, I will demonstrate that unique optical effects can be achieved by coupling multiple nanoscale vortices into complex structures resembling multiple-gear transmissions. Furthermore, it will be revealed that some previously observed strong light focusing effects can be attributed to nanostructure-induced formation and manipulation of optical powerflow vortices. Finally, I will discuss application of these new design principles to the development of multi-functional phase-operated plasmonic machinery, including light-trapping and biosensing platforms, SERS substrates, spectrally-tunable directional nanoantennas, strain sensors, waveguides, and metamaterial building blocks.

## 12.2. Energy transport and dissipation in plasmonic materials

The time- and space evolution of the electromagnetic field vectors **E** and **H** can be obtained by solving Maxwell equations with the problem-specific initial and/or boundary conditions:

$$\begin{aligned} \nabla \cdot \mathbf{E} &= \rho/\varepsilon \\ \nabla \cdot \mathbf{H} &= 0 \\ \nabla \times \mathbf{E} &= -\mu \cdot \partial \mathbf{H}/\partial t \\ \nabla \times \mathbf{H} &= \mathbf{J} + \varepsilon \cdot \partial \mathbf{E}/\partial t \end{aligned} \quad (12.1)$$

However, it is "*more physically satisfying and visually appealing – as well as pedagogically more effective*" [23] to consider the evolution of the dc component of the power density flux, which is given by the time-averaged Poynting vector **S**:

$$\mathbf{S} = 1/2 \, \mathrm{Re}[\mathbf{E} \times \mathbf{H}^*] . \quad (12.2)$$

Eq. 12.1 has a form applicable in the case of an isotropic linear medium, with $\varepsilon$ and $\mu$ being permittivity and permeability, and $\rho$ and **J** - charge and current density, respectively. The Poynting vector specifies the magnitude and direction of the energy flow, and the integral of the power flux through a closed surface gives the rate of the energy loss from the surrounded volume by leakage through its surface. Flow of the electromagnetic energy through absorbing media (such as metals) is accompanied by its irreversible dissipation, which represents a major bottleneck in the engineering of high-efficiency plasmonic devices.

The energy carried by the electromagnetic wave through a dissipative medium is stored partially in the electromagnetic field and partially in the excited charge oscillations in the host medium. In the frame of the semi-classical Drude-Lorentz-Sommerfeld approach, electromagnetic properties of materials with high concentration of free charge carriers that move against the background of positive ion cores can be captured within the well-known expression for the Drude dielectric permittivity. Assuming the excitation by a harmonic field with $\exp(-i\omega t)$ time dependence, the Drude permittivity function derived from the equation of motion of carriers with charge $e$, mass $m$, density $n$, and collision frequency $\gamma$ takes the following complex form:

$$\varepsilon(\omega) = \varepsilon_r(\omega) + i\varepsilon_i(\omega) = 1 - \frac{\omega_p^2}{\omega^2 + \gamma^2} + i\frac{\omega_p^2 \gamma}{\omega(\omega^2 + \gamma^2)}, \qquad (12.3)$$

where $\omega$ is the light angular frequency, $\omega_p = \sqrt{ne^2/\varepsilon_0 m}$ is the plasma frequency, and the effects of interband transitions and quantum size effects for particles much smaller than the carrier mean free path are neglected. Accordingly, the energy density of a harmonic field propagating through a non-magnetic, dispersive and absorptive medium characterized by the Drude permittivity can be written as [37, 51, 52]:

$$W = \varepsilon_0/4 \cdot (\varepsilon_r + 2\omega\varepsilon_i/\gamma) \cdot |\mathbf{E}|^2 + \mu_0/4 \cdot |\mathbf{H}|^2. \qquad (12.4)$$

The energy conservation law in this case reflects the rate of the energy change within the metal volume $V$ due to leakage across its surface $\sigma$ as well as due to volume dissipation:

$$-\int_V \dot{W} dV = \int_\sigma \mathbf{S} d\sigma + \frac{1}{2}\int_V \omega\varepsilon_0\varepsilon_i |\mathbf{E}|^2 dV. \qquad (12.5)$$

One rarely-discussed yet important consequence of Eqs. 12.4-12.5 is that while a plasmonic nanostructure may provide light concentration in the volume below the diffraction limit in the host medium, the effective mode volume – which accounts for the portion of the electric field energy stored inside the metal – can in principle exceed this limit [37]. Furthermore, the energy dissipation rate grows with the increase of the imaginary part of permittivity, and also with the amount of energy stored in the electric rather than magnetic field. Typically, for electromagnetic fields confined in the vicinity of sub-wavelength plasmonic nanostructures, magnetic energy is just a tiny portion of the electric energy. This results in the fast energy dissipation and low degree of temporal coherence of localized SP modes [53]. Typical approaches to alleviate or compensate for dissipation losses include loss compensation with gain [4, 54, 55] – which however may require high pump rates and complicates local heat management in sub-wavelength plasmonic com-

ponents – and the search for new materials with high charge carriers mobility yet reduced dissipative losses [56-59]. In the following, I will discuss an alternative approach to reducing the dissipative losses by modifying the pattern of the nanoscale powerflow through plasmonic nanostructures.

## 12.3. How can a particle absorb more than the light incident on it?

The above question was famously posed and answered by Craig Bohren [23] to explain the physics of the strong energy concentration on metal nanoparticles by studying the modification of the optical powerflow in the electromagnetic field due to the presence of a single nanoparticle. The Poynting vector field lines – which otherwise would have been parallel lines along the propagation direction of the plane wave – become drastically distorted in the vicinity of the particle at the frequencies corresponding to the real parts of the particle eigenmodes frequencies [23, 60, 61]. This effect is illustrated in Fig. 12.4a, which shows the optical powerflow behavior at the wavelength of the dipole LSP mode in an aluminum nanosphere [23]. Far from the sphere, the field lines are still parallel, but they strongly converge to the sphere surface in close vicinity to the particle, increasing the effective particle cross-section, as 'seen' by light. This ability of the particle to harvest light from the area much larger that the particle cross-section translates into the strong field concentration on the particle surface (see inset to Fig. 12.4a). It was initially assumed that away from the LSP resonances, the Poynting vector field lines just bend around the particle, resulting in the absence of the strong field enhancement off resonance (Fig. 12.4b).

In general, the ability of the particle to scatter and absorb light can be characterized by its scattering/absorption cross sections, or scattering/absorption efficiencies (cross-sections normalized to the particle volume). For a nanosphere of radius $a$ and complex-valued permittivity $\varepsilon$ immersed in air, these efficiencies can be found by analytically solving the Maxwell equations in the quasi-static approximation, and have the following form [23]:

$$Q_{abs} \approx 4\left(\frac{2\pi a}{\lambda}\right)\text{Im}\left(\frac{\varepsilon-1}{\varepsilon+2}\right), \quad Q_{sca} \approx \frac{8}{3}\left(\frac{2\pi a}{\lambda}\right)^4 \left|\frac{\varepsilon-1}{\varepsilon+2}\right|^2. \qquad (12.6)$$

Not surprisingly, both, the scattering and the absorption cross-sections grow with increase of the sphere size (although the scattering efficiency clearly prevails for larger particles), and their values resonantly peak at the frequency where $\varepsilon_r = -2$. A somewhat less-expected result is that not only the scattering efficiency and localized field intensity (see e.g. [62]) but also the absorption efficiency is inversely proportional to the $\varepsilon_i$ value at the resonant frequency.

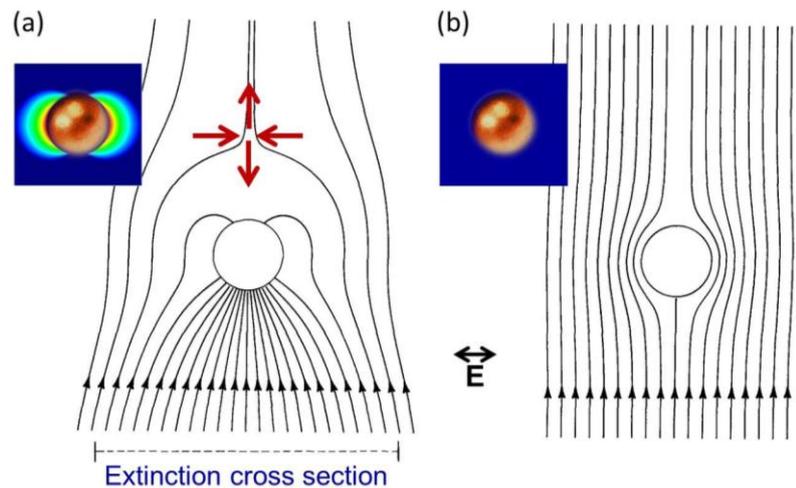

Fig. 12.4. Optical powerflow underlying the dipole LSP resonance in the plane wave scattering by a metal nanoparticle. Field lines of the Poynting vector in the E-plane around an aluminum nanosphere illuminated by a plane wave at the frequency of its dipole LSP resonance (a) and away from it (b) (adapted with permission from [23] ©AIP ). The insets show the corresponding electric field intensity distributions around the particle.

The most striking feature of the on-resonance powerflow pattern shown in Fig. 12.4a is the reversal of the energy flow in the particle shadow area, with a portion of energy entering the particle from behind. This powerflow reversal is driven by the formation of a topological feature in the Poynting vector field - a saddle node. The direction of the powerflow around the saddle node is highlighted with red arrows in Fig. 12.4a. A presence of another saddle node inside the particle volume drives the exchange of the electromagnetic energy between the E- and H-planes. As the energy re-circulation occurs through the volume of the metal – where energy dissipation takes place (12.5) – the LSP resonance is accompanied by high absorption losses. The saddle nodes observed in Fig. 12.4a are the local topological features in the Poynting vector field at which the intensity of the time-averaged Poynting vector is zero, and thus its phase cannot be defined [63-67]. The powerflow lines around the saddle points are hyperbolas (see Fig. 12.4a). It has previously been observed that the decrease of the material dissipation losses results in even more complex picture of the powerflow through the particle [61, 68]. This effect may explain the inverse dependence of the absorption efficiency on the imaginary part of the dielectric function due to more efficient light re-circulation through the metal volume.

The scattering efficiency as well as the intensity of the local electric field concentrated on the particle surface can be further enhanced by combining the individual particles into ordered arrays and making use of the constructive interference of the scattered partial fields. The scattering, absorption and emission spectra of such arrays exhibit sharp resonant peaks [20, 41, 42, 46, 69-71] corresponding to the well-known Rayleigh anomalies associated with the opening of new array diffraction orders. For discrete values of array periods proportional to the wavelength of the incident field, the dipoles induced in individual particles oscillate in phase, inducing coherent correlations across the array. This collective effect – known as superradiance – was initially predicted by Dicke for a system of two-level atoms [72] and results in the generation of a signal proportional to the square of the number of dipoles in the directions of their constructive interference.

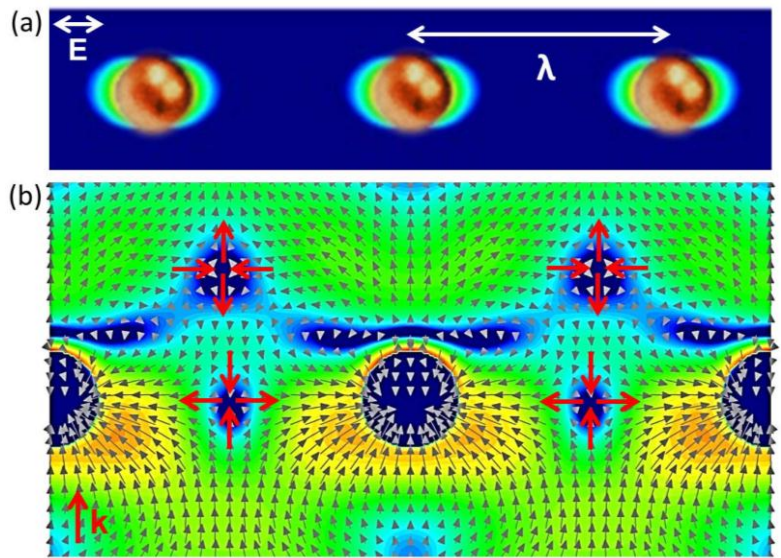

Fig. 12.5. Energy flow underlying the Dicke superradience effect in a periodic nanoparticle array. (a) Electric field intensity distribution in the E-plane of the periodic chain of Au nanoparticles separated by distances close to the wavelength of the illuminating plane wave. (b) Power-flow pattern (small arrows) and the spatial intensity map of the Poynting vector in the plane cutting through the centers of the nanoparticles in the chain. The direction of the powerflow around the additional array-induced saddle nodes is highlighted with red arrows.

A complex picture of energy flow through a periodic chain of gold nanoparticles at the frequency of the chain superradiant collective mode resulting from the spectral overlap of a particle dipole LSP resonance and a chain Rayleigh anomaly is illustrated in Fig. 12.5b. It can be seen that – similarly to the case of individual nanoparticles – the field lines strongly deflect towards particles, and the energy is re-circulated through their metal volumes. Furthermore, formation of a collective mode is associated with appearance of additional topological features around the particle chain (the direction of powerflow around additional saddle nodes is indicated with red arrows in Fig. 12.5b). The net result of this effect is a more efficient light trapping and re-circulation through the particle chain, which increases its col-lective scattering, absorption, and field enhancement efficiency over that of a sum of efficiencies of individual nanoparticles.

## 12.4. Nanoparticle-generated optical tornadoes

Saddle nodes are not the only type of topological features that can emerge as a result of the interference of several waves (or – in the alternative hydrodynamic picture – collision of several flows). Perhaps the most intriguing flow features are flow vortices – singular nodes of the Poynting vector (points in 2D and lines in 3D) – that are the centers of circulating powerflow (i.e., the field lines surrounding a vortex are circles) [63, 65, 73]. These optical analogs of whirlpools and tornadoes can be created by superposition of three (or more) waves, e.g., plane waves, partially scattered fields, pulses, or fields generated by point sources. Understanding the origins and exploiting flow effects associated with these topological features have been proven to be of high relevance for many branches of physics, including hydrodynamics, acoustics, quantum physics, and singular optics [73, 74].

Formation of vortices is known to significantly affect the properties of superfluids, superconductors, and optical components; however, without proper engineering their effect can be detrimental. In particular, electromagnetic vortices form in the interference field of the waves propagating through random, turbulent and chaotic media and are manifested as dark points in the speckle fields. However, their presence causes trouble in the measurement and interpretation of scattered fields for remote sensing applications. Formation of vortices in the interference field of light scattered by sub-wavelength slits can also be detrimental as it results in the phenomenon of the frustrated transmission [66]. Likewise, the motion of 'Abrikosov vortices' – nanoscale tubes of magnetic flux that form inside superconductors upon exposure to magnetic fields – decreases the current-carrying capability of superconductors and requires vortex immobilization ( or 'pinning') by the material artificial defects [75, 76]. To pin vortices effectively, in-depth understanding of the pinning strength of individual defects as well as the collective effects of many defects interacting with many vortices is required.

In a perfect analogy, developing a general design strategy of pinning optical vortices to properly-designed plasmonic nanostructures would offer an unprecedented degree of control over light trapping in sub-wavelength volumes and in dynamical manipulation of nanoscale powerflow patterns. We already saw in the previous sections that resonant scattering of light from noble-metal nanoparticles creates a complex spatial distribution of the time-averaged Poynting vector in their near-field. An even more interesting nanoscale powerflow features can be observed in at the frequencies just around the LSP resonant frequency in the nanoparticle, as illustrated in Figures 12.6 and 12.7. Fig. 12.6a shows frequency spectra of the electric and magnetic field intensities generated on the particle surface by a plane wave propagating along the z-axis and linearly polarized along the x-axis.

The intensity values are probed at intersections of the particle equatorial plane with the E- and H-planes (see inset to Fig. 12.6a). The excitation of the dipole LSP resonance in the nanoparticle is marked by the pronounced peaks in the E-field intensity spectra (with the E-plane intensity value being an order of magnitude higher). The magnetic field intensity probed in the H-plane also features a weak resonant feature in the same frequency range.

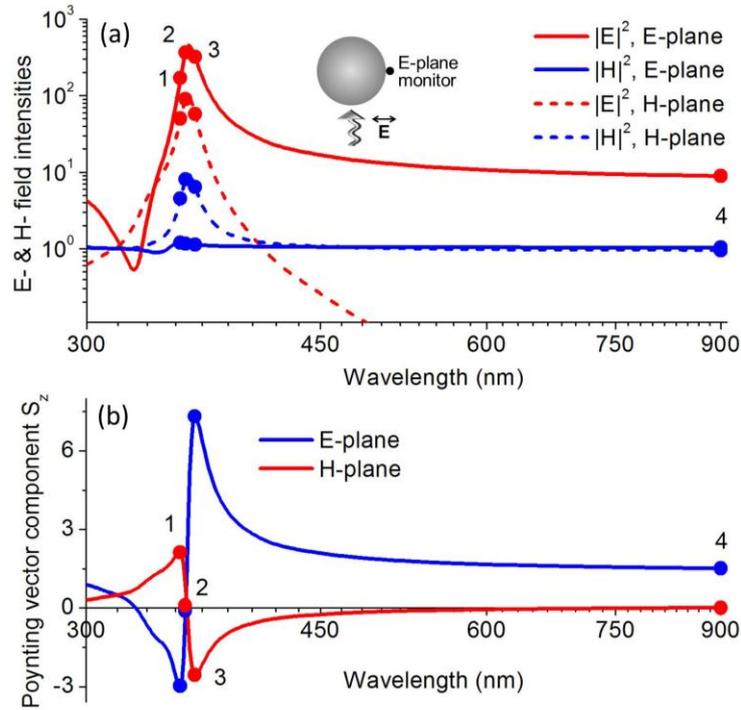

Fig. 12.6. Optical powerflow around a plasmonic nanoparticle: off-resonance effects. Wavelength spectra of the electromagnetic field intensities (a) and the energy flow direction (b) around a 30nm-diameter Ag nanosphere in a wide spectral window enclosing the particle LSP resonance. The particle is illuminated by a linearly-polarized plane wave propagating upwards along the z-axis. The inset in 6a shows the spatial position of the field monitor in the E-plane (the H-plane monitor is located at the same position in the H plane). The dots indicate the discrete wavelengths at which the spatial distributions of the field intensity and powerflow patterns are calculated (see Fig. 12.7).

To reveal the direction and density of the optical power flux around the nanoparticle on and off its LSP resonance, the frequency spectra of the z-component of the Poynting vector (pointing in the direction of the plane wave propagation) are plotted in Fig. 12.6b. The Poynting vector (12.2) is a 3D real-valued vector field, which in the Cartesian coordinate system is defined by three orthogonal components:

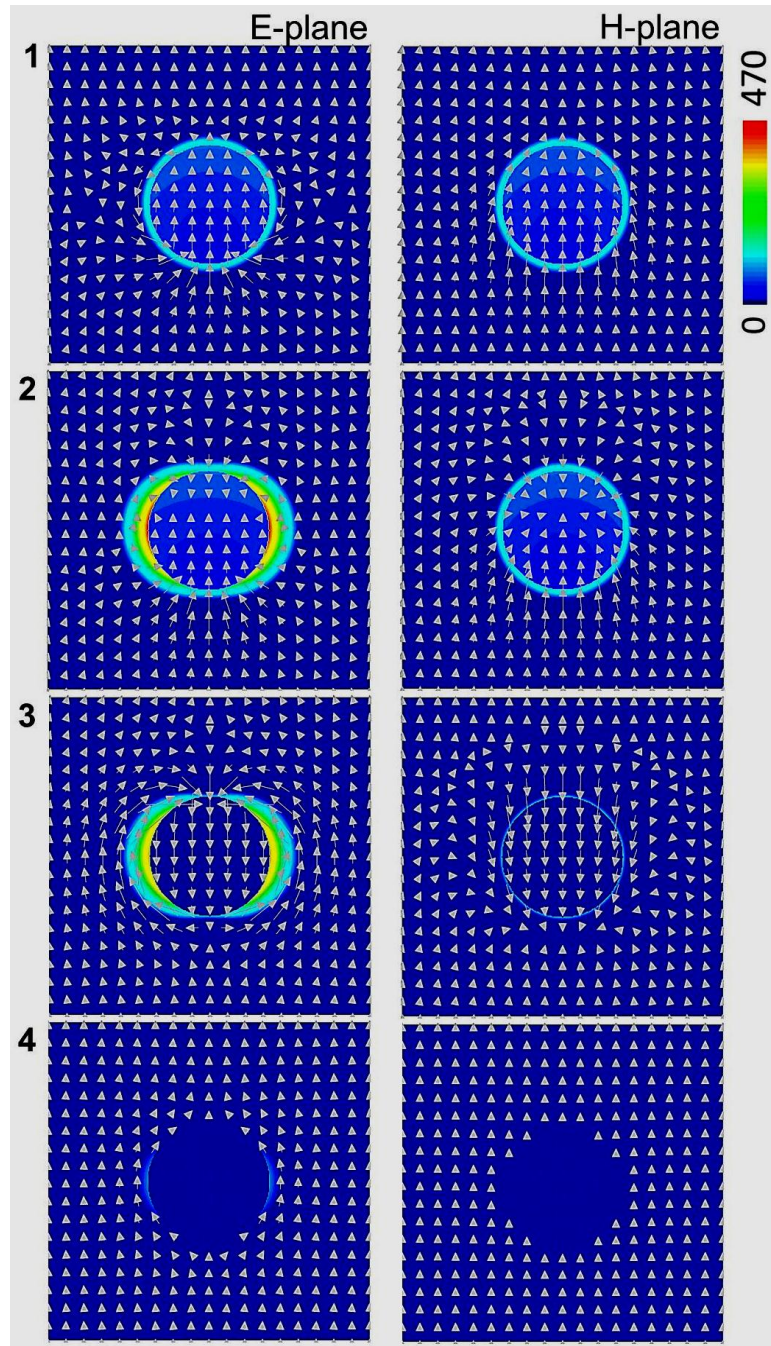

Fig. 12.7. Optical powerflow (arrows) and E-field intensity patterns on a 30nm-diam Ag sphere at the frequencies around its dipole LSP resonance (positions 1-4 are marked in Fig. 12.6).

$\mathbf{S} = \{S_x, S_y, S_z\}$. The plots in Fig. 12.6b show that the direction of the powerflow around the particle changes abruptly from forward (characterized by positive $S_z$ values) to backward (negative $S_z$) at the wavelength of the LSP resonance. The phase change by π at a resonance is a manifestation of the delay between the action of the driving force and the oscillator response. It is a characteristic feature of any resonant structure driven by the external field, including classical forced oscillators [77, 78]. However, the spatial distributions of the Poynting vector field around the particle on and off resonance shown in Fig. 12.7 reveal an intricate picture of local powerflow. In particular, a pair of saddle nodes can be observed in the on-resonance powerflow pattern 2, while at the frequencies just around the LSP resonance, the powerflow forms pairs of coupled counter-rotating vortices (patterns 1 and 3 in Fig. 12.7). Only far away from the resonant frequency (pattern 4) the flow resembles the one discussed by Bohren (compare to Fig. 12.4b). Hereafter, the arrows point into the direction of the local powerflow, and the length of each arrow is proportional to the local value of the Poynting vector amplitude.

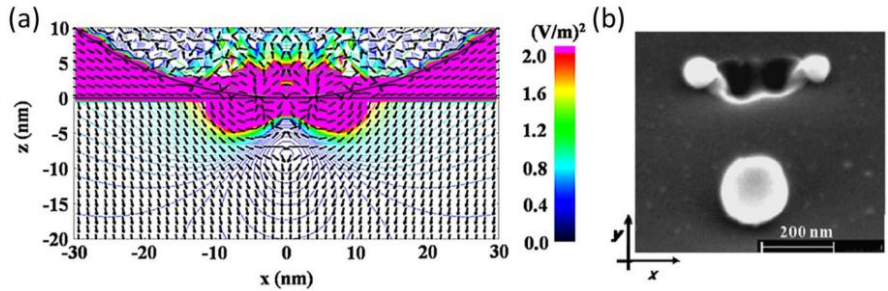

Fig. 12.8. Near-field nanopatterning with plasmonic particles. (a) Optical powerflow (arrows) and the electric field intensity distribution (color) around a 100nm-diam Au nanosphere deposited on a silicon substrate (reproduced with permission from [79], ©Springer). (b) SEM image of the nanohole fabricated by the particle-enhanced near-field under the laser illumination (reproduced with permission from [80], ©Elsevier).

First predicted theoretically several years ago [68, 81], the effect of formation of optical whirlpools in individual nanoparticles initially did not attract a lot of attention because it does not translate into efficient field enhancement (see Fig. 12.6a). However, the situation changes when the particles are deposited on a high-refractive-index dielectric substrate as illustrated in Fig. 12.8 [79]. In particular, resonant illumination of an Au particle by a focused laser beam may cause formation of coupled counter-rotating vortices at the interface between the particle and the substrate (shown in Fig. 12.8a). The strong light re-circulation through a nanoscale area underneath the particle creates localized temperature gradient in the substrate material, and can be used for efficient nanohole processing (Fig. 12.8b) [80].

## 12.5. Molding the river of light in vortex nanogear transmissions

As shown in the previous sections, nanostructure-induced interference of the evanescent-field components of the electromagnetic field may generate complex spatial distributions of the time-averaged Poynting vector in the near-field region, including creation of nanoscale vortices of optical powerflow. This mechanism of light trapping offers a new way to efficiently focus and store light within nanoscale volumes. However, it calls for more work to formulate rational strategies for vortex-pinning nanostructures engineering and to predict new physical effects. It has long been realized that variations of amplitude and phase of interacting *propagating waves* can be used to create and control *free-space vortex fields*. Experimental techniques for the free-space vortices generation include the use of "forked" holograms, lenses and spiral phase plates as well as their recently proposed plasmonic analogs [82-84]. Complex vortex topologies such as loops and knots can be created by superposing waves with correctly-weighted amplitudes and phases [85, 86]. A similar theoretical framework for generation and manipulation of the powerflow vortices via the interference of evanescent waves in plasmonic nanostructures is still under development [49, 67, 87], and below I will review several successful realizations of the vortex-pinning nanostructures.

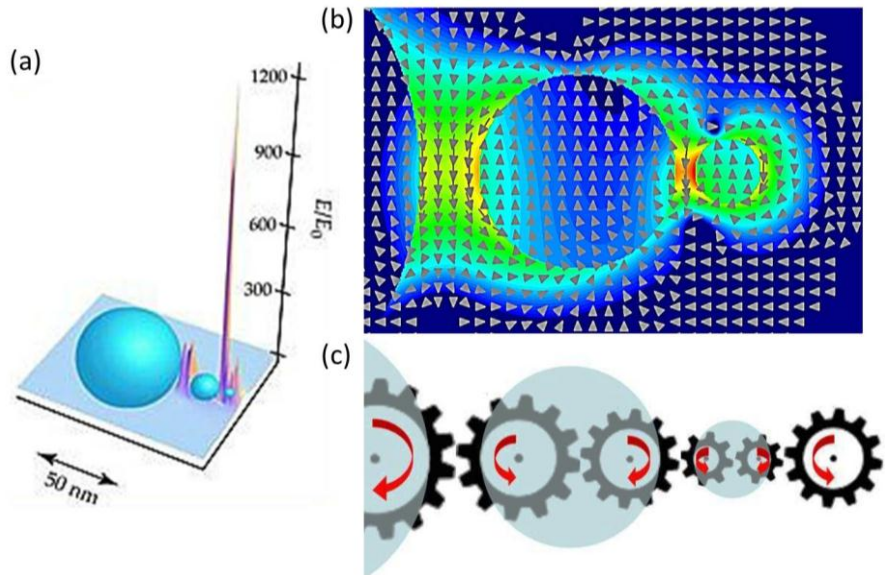

Fig. 12.9. Silver nanolens as a vortex nanogear tansmission. (a) Schematic of a nanolens and an electric field intensity distribution generated by an incoming plane wave (reproduced with permission from [88] ©APS). (b) Optical powerflow pattern underlying the nanolens focusing mechanism. (c) Schematic of the nanogear transmission generated in the nanolens. Power flux in each nanogear is looped through nanoparticles (adapted with permission from [67] ©RSC).

The first example to be discussed is a celebrated silver nanolens design for efficient light focusing in the gap between the particles arranged into a showman-like structure (shown in Fig. 12.9a). This configuration was theoretically introduced over a decade ago by Li, Stockman and Bergman [89], and has generated over 300 follow-up papers since publication. However, the origin of the observed phenomenon has not been understood until recently [67], and can be revealed through the concept of coupled-vortex formation. To illustrate the underlying physical effects of light focusing in the nanolens, in Fig. 12.9b the on-resonance optical powerflow pattern is overlapped with the electric field intensity distribution. The structure is illuminated by a linearly-polarized plane wave (propagating from bottom up in Fig. 12.9b). Figure 12.9b reveals a complex picture of powerflow through the nanolens focal point, with the reversal of flow direction at the near-field intensity resonance, which is caused by the formation of coupled counter-rotating vortices inside the nanoparticles. The powerflow pattern in Fig. 12.9b resembles a multiple-gear transmission (schematically illustrated in Fig. 12.9c), which is composed of coupled vortex nanogears made of light and arranged into a linear sequence. To reflect this analogy, the nanostructures engineered to couple counter-rotating optical vortices into transmission-like sequences have been termed *Vortex Nanogear Transmissions* (VNTs) [67, 87]. As seen in Fig. 12.9b, light circulation through VNT drives the increased backward powerflow through the inter-particle gaps, resulting in the dramatic local field enhancement.

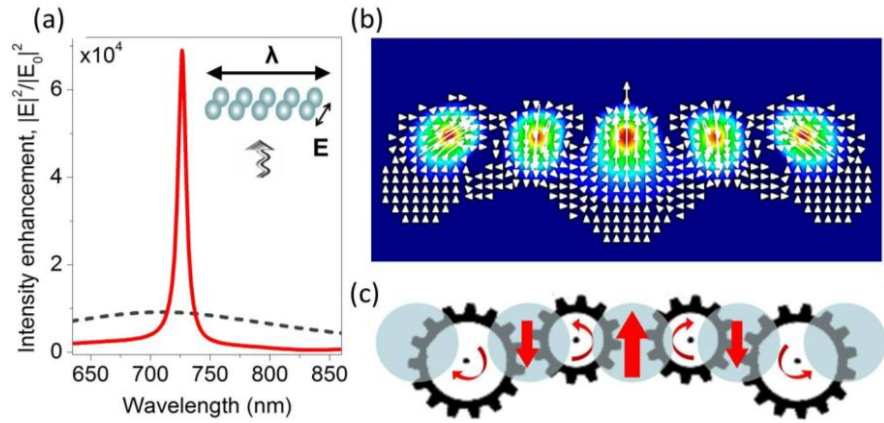

Fig. 12.10. Vortex nanogear transmission re-circulating optical energy outside of the nanoparticles metal volume. (a) Wavelength spectrum of the electric field intensity enhancement in the center of a chain of Ag nanosphere dimers of 50nm sphere radii, 3nm intra-dimer gaps and 10nm inter-dimer gaps in water, as shown in the inset. The corresponding spectrum of a single dimer is shown as dashed line. The structures are illuminated by a plane wave with the electric field polarized along the dimers axes. (b). Poynting vector intensity distribution and powerflow through the chain (plotted in the plane cutting through the centers of the dimer gaps) at its resonant wavelength. (c) Schematic of the VNT generated in the dimer chain. Power flux in each nanogear is looped between the particles (adapted with permission from [67] ©RSC).

The nanolens in Fig. 12.9 re-circulates optical power through the volume of nanoparticles, which results in the strong energy dissipation on every pass through the metal. Accordingly, engineering VNTs capable of weaving the powerflow *away from* the volume of metal nanoparticles would go long way in overcoming the problem of high dissipative losses in plasmonics even without using gain compensation mechanism. Physically, there are no restrictions on creating optical vortices outside of the nanostructure. Vortices – as well as saddle nodes – are characterized by the zero energy absorption ($\nabla \cdot \mathbf{S}(\mathbf{r}) = 0$) and thus can exist in the non-absorbing medium [63-67]. Indeed, various configurations of the VNTs capable of folding light into nanogears threaded through the inter-particle gaps have been predicted theoretically [49, 67, 87] and subsequently realized experimentally [87]. One such design is shown in the inset to Fig. 12.10a, and consists of a regular linear chain of identical Ag nanoparticle dimers separated by sub-wavelength gaps [67, 87]. This VNT can generate localized light intensity in the gap of the central dimer, which significantly exceeds that in a stand-alone dimer (Fig. 12.10a).

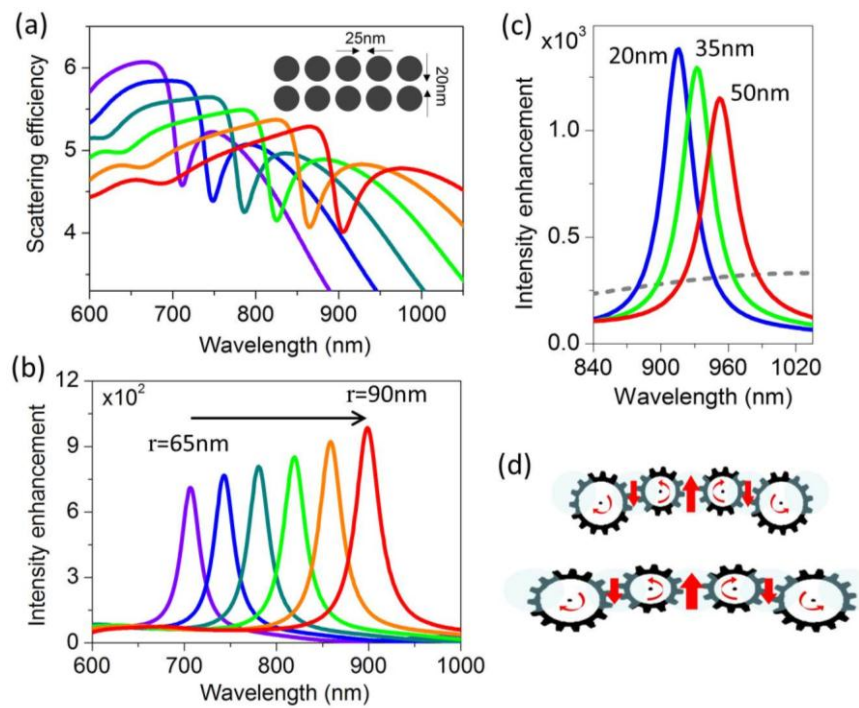

Fig. 12.11. Rainbow trapping in spectrally-tunable VNTs. Red-shifting of the spectral features in the scattering efficiency (a) and the localized electric field intensity (b) of the linear VNT shown in the inset. Au nanoparticles radii are varied from 65nm to 90nm with 5nm increment. (c) Tuning of the VNT spectrum via the increase of the inter-dimer gaps, which amounts to 'stretching' the vortex nanogears, as schematically illustrated in (d).

Note that the reduced powerflow through the metal results in the significant narrowing of the resonance linewidth, reflecting the increase of the quality (Q) factor of the collective VNT mode over that of a single dimer LSP mode. The Q-factor is inversely proportional to the energy loss rate of the mode. It can be interpreted as the number of field oscillations that occur coherently, during which the mode is able to sustain its phase and accumulate energy. Accordingly, local fields can be enhanced by a factor of Q (as seen in Fig. 12.10a). Other important characteristics of the mode interaction with its material environment also scale with Q, including spontaneous and stimulated emission rates, absorption rates, and the strength of resonant Coulomb interaction between distant charges [37, 90, 91].

Metal nanoparticle structures capable of pinning optical vortices and combining them into nanogear transmissions can be easily fabricated by standard techniques such as electron beam lithography [87, 92] or template-assisted self-assembly [13]. Furthermore, their spectral properties can be tuned in a wide frequency range by varying the nanoparticle sizes as well as the separations between adjacent dimers (see Fig. 12.11). In particular, the possibility of tuning the resonant frequency of a linear VNT by stretching has the potential for developing mechanically- and optofluidically-tunable devices. Note that the optical spectrum of a single nanodimer can also be tuned by pulling the two nanoparticles away from each other by stretching the substrate. However, the increase of the gap size has a severe detrimental effect on the dimer near-field intensity. To the contrary, the spectra of linear VNTs can be tuned by stretching the substrate in the direction along the VNT long axis (see Fig. 12.11c,d). This deformation does not result in the increase of the intra-dimer gaps (and for sufficient deformations can even shrink the gaps). This is reflected in the very slight variations in the peak intensity for a significant (over 100%) VNT stretching (Fig. 12.11c).

Although the position of each optical vortex pinned to the nanostructure is sensitive to changes in the structure geometry, vortices are structurally stable and typically only continuously migrate upon small perturbations of field parameters [63, 65]. This opens the way – exploited in the designs shown in Fig. 12.11 – to continuously tune the VNT characteristics in a controllable fashion by changing the nanostructure design. On the other hand, large perturbations of the controllable external parameters (such as e.g. wavelength of incident field) can be introduced to force vortices of opposite sign either to approach and annihilate, or to nucleate as a pair. The above effects form the basis for realizing active spatio-temporal control of powerflow and the field intensity redistribution on the nanoscale, as illustrated in Fig. 12.12.

In particular, Fig. 12.12 shows a design of a 'vortex nanogate', in which optical powerflow though an Au nanodimer gap coupled to a dielectric microsphere (as shown in Fig. 12.3a) can be dynamically switched on/off and even reversed by tuning the frequency of the excitation [49]. As seen in Fig. 12.12a, under the resonant excitation by the plane wave, pairs of coupled counter-rotating optical vortices form both inside the microsphere and around the Au nanodimer gap. Their combined effect yields enhanced backward powerflow through the dimer gap (Fig. 12.12a). With the wavelength shift, the vortices migrate away from the nanodimer gap area, while oppositely-rotating vortices form there and help to drive the enhanced forward powerflow through the nanogate as shown in Fig. 12.12(c). A schematic of various arrangements of vortex nanogear configurations corresponding to different regimes of the nanogate operation (i.e., 'open up' or 'open down') are schematically visualized in Figs. 12.12b,d. The proposed strategy of reconfiguring powerflow through plasmonic nanostructures opens the way to developing chip-integrated plasmonic and optoplasmonic switching architectures, which is crucial for implementation of quantum information nanocircuits. It should also be noted that vortex-driven enhanced or blocked powerflow through sub-wavelength gaps was revealed [66] to be behind the phenomenon of extraordinary transmission through nanohole arrays [93].

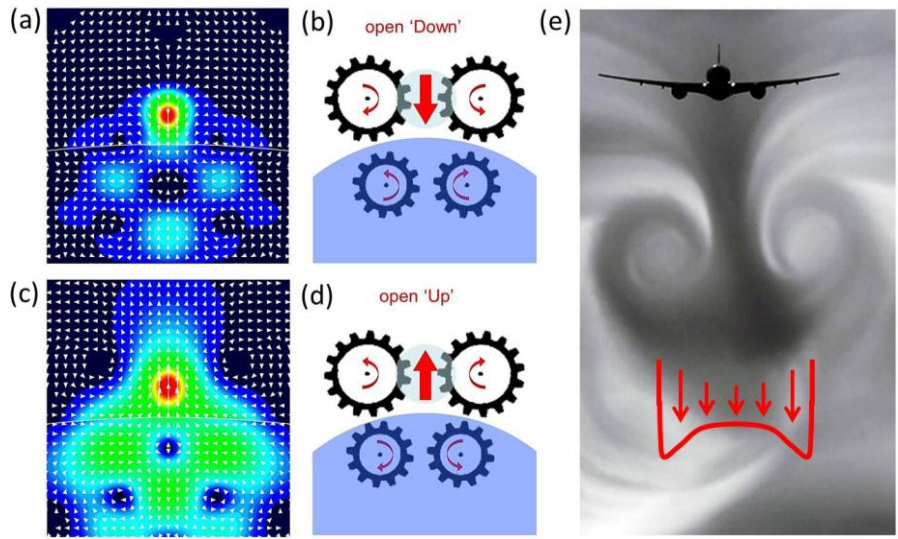

Fig. 12.12. Operation of the optoplasmonic vortex nanogate. (a,c) Poynting vector intensity maps and the optical powerflow through the gap of a microsphere-coupled Au nanodimer at the frequencies around the resonance of the structure shown in Fig. 12.3a. Spatial maps are shown in the plane cutting through the dimer gap center. (b,d) Schematics of the vortex-operated nanogate in the 'open Down' and 'open Up' positions (reproduced with permission from [49] ©OSA). (e) Coupled vortices forming behind a flying airplane (source: [94]), with a vortex-driven downwash velocity profile shown as the inset.

Overall, a rational design strategy based on creation and coupling of areas of circulating powerflow offers a new level of control over tailoring optical properties of plasmonic nanocircuits, possibly providing a better insight into their behavior than traditional approaches based on antenna design concepts. Furthermore, owing to the similarities between optical, hydrodynamic and superfluidic vortices, many concepts can be directly borrowed or adapted from other branches of phys-ics. In particular, enhanced vortex-driven powerflow through the dimer gap shown in Figs. 12.12a,c remarkably resembles a picture of coupled vortex flows that combine to provide a downwash velocity profile behind a flying airplane (Fig. 12.12e). The fundamental mechanisms underlying this intuitive picture can be revealed by using the mathematical isomorphism between the hydrodynamic equations and the electromagnetic wave equations separated into phase and amplitude variables, as will be discussed in the following section.

## 12.6. Hydrodynamics of light flow in plasmonic nanostructures

Maxwell famously pointed out defending his theory of molecular vortices – which aimed at explaining electromagnetic phenomena mechanically in terms of the interacting vortices in the imaginary fluid ether – that "*it is a good thing to have two ways of looking at a subject and to admit that there are two ways of looking at it*" [95]. It has long been recognized that hydrodynamics offers an alternative way of looking at electromagnetic waves propagation even in a very general case of dispersive, lossy and nonlinear media [96-101]. For example, a nonlinear Schrödinger equation that governs the amplitude of the wave trains propagating through nonlinear medium in a paraxial approximation can be re-cast in a hydrodynamic form via the Madelung transformation $\psi(\mathbf{r},t) = \sqrt{\rho(\mathbf{r},t)} \exp\{i\Phi(\mathbf{r},t)\}$:

$$\begin{aligned}&\frac{\partial \rho}{\partial \tau} + \nabla'(\rho \mathbf{v}) = 0 \\ &\frac{\partial \mathbf{v}}{\partial \tau} + (\mathbf{v} \cdot \nabla')\mathbf{v} = -\frac{1}{\rho}\nabla'\left(\frac{\rho^2}{2}\right) + \nabla' Q\end{aligned}. \quad (12.6)$$

Eqs. 12.6 have a well-known form of the Navier-Stokes equations of the fluid dynamics for inviscid, compressible and irrotational flow, with the field intensity $|\psi|^2$ regarded as the 'photon fluid' (PF) density $\rho$, and the phase gradient $\nabla'\Phi$ as the PF velocity $\mathbf{v}$. Note that to arrive to the familiar form of (12.6), the time is treated as the third spatial coordinate: $\nabla' = \hat{\mathbf{x}}\partial/\partial x + \hat{\mathbf{y}}\partial/\partial y + \hat{\mathbf{t}}\partial/\partial t$, and the space coordinate in the propagation direction – as a generalized time coordinate $\tau$: $\partial/\partial z \equiv \partial/\partial \tau$. The first equation in (12.6) has the same form as the continuity (mass conservation) equation of the fluid dynamics. The right hand side of the momentum conservation equation (second line in (12.6)), however, features not only the term analogous to the fluid pressure but also a term $Q = \nabla'^2\sqrt{\rho}/2\sqrt{\rho}$ known as 'quantum pressure', which has no analog in hydrodynamics.

The availability of the hydrodynamic form (12.6) of the nonlinear Schrödinger equation helps to draw parallels between the properties of lasers, superfluids and superconductors. A similar, but rarely invoked analogy can be drawn between the electromagnetic Maxwell equations and the hydrodynamics equations. In particular, the Helmholtz wave equation for a monochromatic plane wave propagating in a piece-wise homogeneous linear nonmagnetic medium with complex permittivity $\varepsilon(\mathbf{r}) = \varepsilon_r(\mathbf{r}) + i\varepsilon_i(\mathbf{r})$ can be cast in a from similar to that of the Schrödinger equation with the external potential $V(\mathbf{r}) = k_0^2/2(1 - \varepsilon_r(\mathbf{r}))$ ($V(\mathbf{r}) = 0$ in the free space) and the total energy $E_0 = k_0^2/2$ [78]. Subsequent application of the Madelung transformation $\mathbf{E}(\mathbf{r}) = \sum_{m=1}^{3} \hat{e}_m U_m(\mathbf{r}) \exp\{i(\Phi(\mathbf{r}) - \omega t)\}$ brings the wave equation to the hydrodynamic form for the steady flow ($\partial \rho/\partial t = 0$) of 'photon fluid' in the presence of sources and sinks $\alpha(\mathbf{r}) = -k_0^2 \varepsilon_i(\mathbf{r})$ [67, 102]:

$$\begin{aligned} \nabla(\rho(\mathbf{r})\mathbf{v}(\mathbf{r})) &= \alpha(\mathbf{r})\rho(\mathbf{r}) \\ (\mathbf{v}(\mathbf{r}) \cdot \nabla)\mathbf{v}(\mathbf{r}) &= -\nabla(V(\mathbf{r}) + Q(\mathbf{r})) \end{aligned} \quad (12.7)$$

Similarly to Eq. 12.6, here the intensity plays the role of a 'photon fluid' density $\rho(\mathbf{r}) = \mathbf{U}(\mathbf{r}) \cdot \mathbf{U}(\mathbf{r})$, and the phase gradient – the role of the fluid velocity, $\mathbf{v} = \nabla \Phi(\mathbf{r})$. However, the local pressure term does not appear in the absence of non-linearity, and the internal quantum potential has the following form: $Q(\mathbf{r}) = 1/(8\rho(\mathbf{r})) \sum_{m=1}^{3} (\nabla \rho_m(\mathbf{r}) \cdot \nabla \rho_m(\mathbf{r})/\rho_m(\mathbf{r})) - \nabla^2 \rho(\mathbf{r})/(4\rho(\mathbf{r}))$, where $\rho_m(\mathbf{r}) = U_m^2(\mathbf{r})$. Accordingly, the optical flux defined by the Poynting vector transforms into the analog of a fluid flux (the momentum density): $\mathbf{S} = 1/(2\mu_0\omega) \cdot \rho(\mathbf{r})\mathbf{v}(\mathbf{r})$. One important difference between conventional fluids and the PF is that photon 'mass' can be created owing to the linear gain $\varepsilon_i(\mathbf{r}) < 0$ and dissipated through material losses $\varepsilon_i(\mathbf{r}) > 0$. The 'photon mass' reduction due to dissipative losses results in the decrease of the PF density (i.e., field intensity).

In the case of PF steady flow, the field patterns are constant in time, yet local accelerations/decelerations of the flow can occur between different parts of the plasmonic nanostructure, governed by the convective term $(\mathbf{v}(\mathbf{r}) \cdot \nabla)\mathbf{v}(\mathbf{r})$ in the momentum conservation equation. This situation is analogous to e.g. local variations of the velocity of the fluid flow passing through pipes of varying diameters. In the structures shown in Figs. 12.10-12.11, these local changes of flow rate are driven by the formation of vortices, with each vortex line inducing a velocity field given by the Biot-Savart formula [103] (compare to the vortex-driven airplane downwash velocity profile shown as the inset of Fig. 12.12e). In particular, the tangential velocity of an optical vortex varies inversely with the distance from its center, and the angular momentum is thus uniform everywhere throughout the vortex-induced circulating flow.

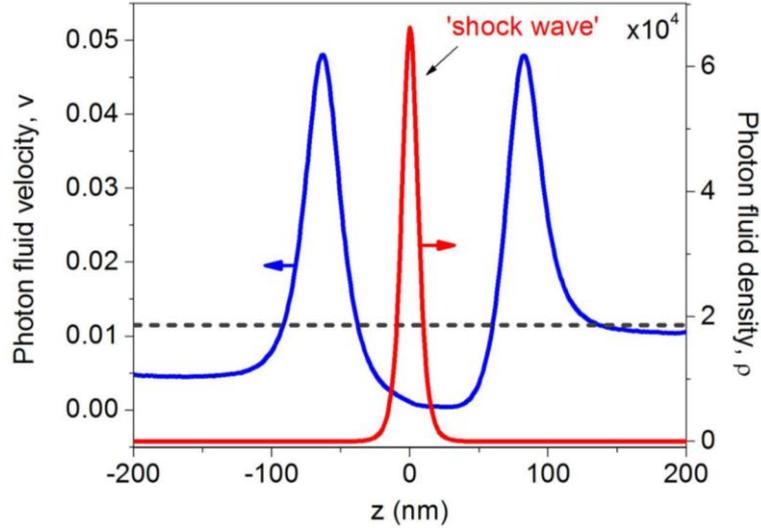

Fig. 12.13. VNT as an analog of a hydraulic pump. Velocity and density of the 'photon fluid' along the line parallel to the incident plane wave propagation direction and passing through the center of the linear VNT in Fig. 2.10. The dash line corresponds to the constant fluid velocity in the absence of the nanostructure. Formation of the 'shock wave' accompanied by the spatially-localized increase of the fluid density in the central dimer gap is observed (reproduced with permission from [67] ©RSC).

A change in the fluid's momentum can generate pressure, and this hydrodynamic effect is utilized in hydraulic pumps and motors, which increase the fluid kinetic energy (angular momentum) and then convert it into usable pressure energy. Accordingly, the problem of wave scattering by the VNT in Fig. 12.10 can be cast in new light by invoking the fluid dynamic analogy. Fig. 12.13 follows the evolution of the PF velocity and density along the $z$-axis, which passes from the bottom up through the gap of the VNT central dimer. At the VNT center, the flows generated by adjacent counter-rotating vortices collide and form a 'shock wave' in the form of a region of high PF density. In the situation depicted in Fig. 12.13, the PF gets convectively accelerated by the potential forces and impacts onto the narrow interparticle gaps of the VNT dimers. The threading of the PF through the gaps leads to a sudden change in the flow regime and results in the dramatic increase of the PF local density driven by the conversion of the PF kinetic energy into pressure energy. In effect, VNTs shown in Fig. 12.10-12.11 operate as plasmonic analogs of hydraulic pumps that exploit the convective acceleration of the PF caused by its circulation through the vortices 'pinned' to the nanostructure to generate localized areas of high PF density [67].

## 12.7. Applications of plasmonically-integrated tornadoes

As demonstrated in the previous sections, nanostructure-generated fine features of the electromagnetic energy flow – which may not be apparent from the near-field intensity patterns – can, however, be controllably manipulated to tailor local properties of the electromagnetic field. The logical next step is the conversion of these theoretical insights into useful practical applications. It is expected that the new approach to light trapping in nanoscale volumes by the generation of optical tornadoes pinned to plasmonic nanostructures may lead to novel device solutions in light generation, harvesting and processing. Here are some specific application areas where this new plasmonic engineering concept offers advantages.

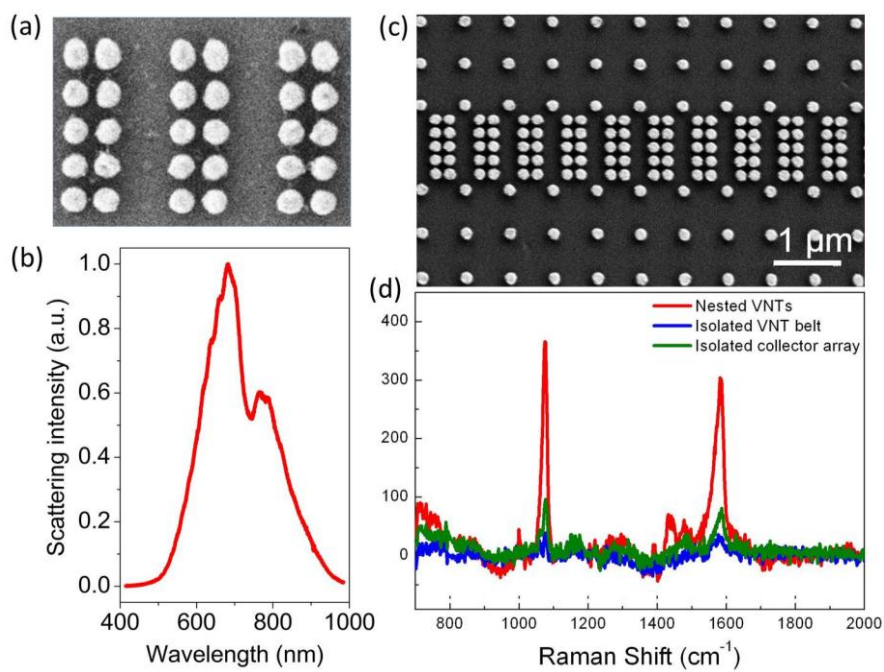

Fig. 12.14. Elastic and inelastic scattering characterization of fabricated VNTs. (a) SEM image of planar Au VNT structures fabricated by e-beam lithography on a quartz substrate (particle diameters 139.2 nm, intra-dimer gaps 30.0nm and inter-dimer gaps 46.4nm). (b) Experimental dark-field scattering spectrum of VNTs shown in (a). (c) SEM image of a SERS platform that combines two mechanisms of field enhancement – Dicke effect in a periodic array and light recirculation through VNTs. (d) Experimental SERS spectra of pMA on the platform shown in (c) (adapted with permission from [87] ©ACS).

The most obvious advantages are envisioned in making use of the VNTs ability to generate high field intensity for Raman and fluorescence spectroscopy. The first surface-enhanced Raman sensing (SERS) platforms have already been successful-ly fabricated by e-beam lithography on planar dielectric substrates (Fig. 12.14a,c). They have been characterized by elastic scattering spectroscopy (Fig. 12.14b), and used to detect Raman spectra of para-mercaptoaniline (pMA) molecules [87] (Fig. 12.14d). To generate a multi-frequency response, different light trapping techniques can be combined within a single platform [7, 49, 69, 104, 105], e.g., Dicke effect in a periodic arrays and light re-circulation through VNTs [87]. Furthermore, it was theoretically predicted that embedding of dipole emitters in VNTs can result in the manifold resonant enhancement of their radiative rates over those of emitters embedded in the gaps of isolated dimers [67]. VNTs can also strongly modify the angular distribution of light emitted by the embedded dipole. This offers opportunities for surface-enhanced fluorescence microscopy and vortex-mediated radiative engineering [67].

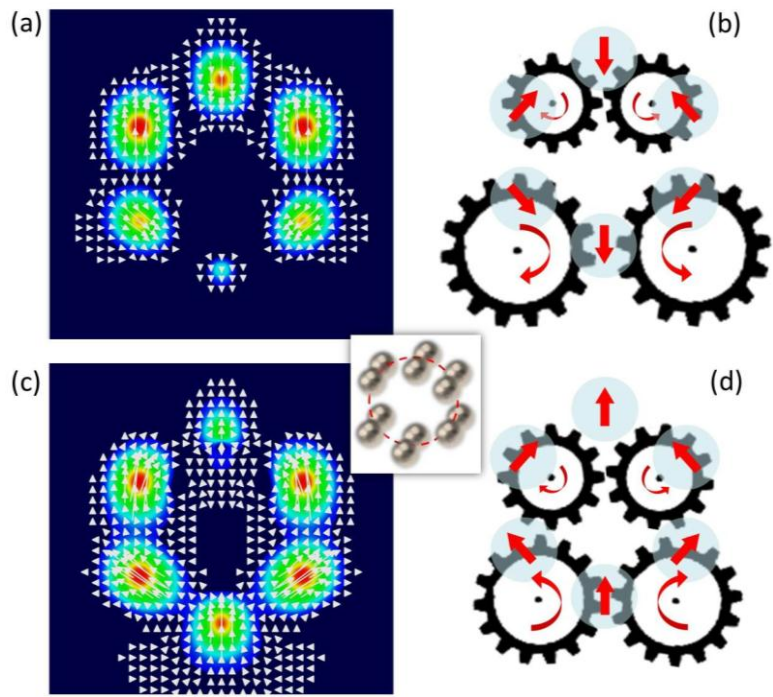

Fig. 12.15. Re-configurable looped nanogear transmission. (a,c) Poynting vector intensity distribution and powerflow pattern through the looped dimer chain shown in the inset at different wavelengths. The inset shows a schematic of a ring-like arrangement of Ag nanoparticle dimers of 50nm radii, 3nm intra-dimer gaps, and 10nm inter-dimer gaps in water. (b,d) Schematics of the coupled and uncoupled VNTs generated by the structure at different resonant wavelengths. Power flux in each nanogear is looped through the gaps between the particles (reproduced with permission from [67] ©RSC).

As already discussed in previous sections, the new way to manipulate powerflow by creating, moving and annihilating nanoscale optical vortices offers tremendous opportunities for intensity switching and energy transfer in plasmonic nanocircuits. Just as mechanical gears and hydrodynamic turbines form the basis of complex machinery, vortex nanogears can be combined into complex reconfigurable networks to enable dynamical light routing. One possible realization of a reconfigurable plasmonic VNT with a footprint of ~320 nm$^2$ is shown in Fig. 12.15 [67]. It consists of six Ag nanodimers arranged into a symmetrical ring-like structure and features several resonances in its optical spectra. The vector fields of the powerflow evaluated at two different resonant wavelengths show drastically different pictures of energy circulation through the nanostructure (Fig. 12.15a,c). In particular, formation of a looped VNT composed of four coupled vortex nanogears is observed in Figs. 12.15c,d. In contrast, when the same structure is illuminated by light of a slightly shifted wavelength, it behaves as a pair of uncoupled two-gear transmissions, thus reversing the powerflow direction as shown in Figs. 12.15b,d.

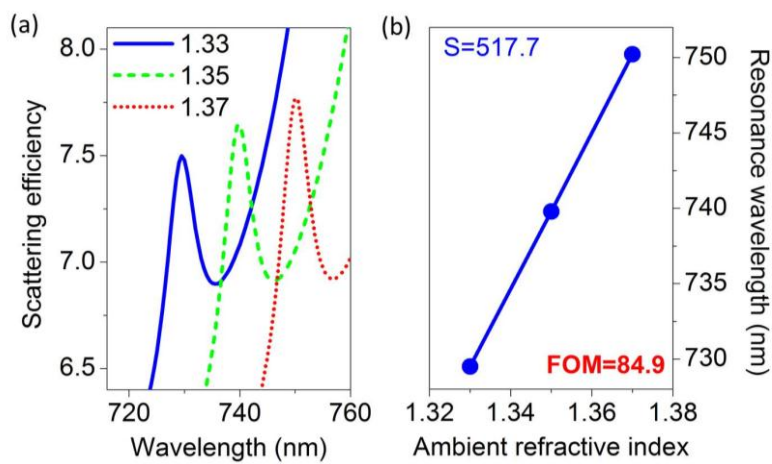

Fig. 12.16. Refractive index sensing with vortex nanogear transmissions. (a) Shift of the far-field scattering efficiency spectrum of the looped VNT in Fig. 12.15 with the change of the refractive index of the ambient medium. (b) The positions of the resonant peaks as a function of the refractive index. The calculated values of the refractive index sensitivity (S) and the corresponding figure of merit (FOM) are also shown.

Another field that can benefit from the hydrodynamics-inspired high-Q plasmonic components is bio(chemical) sensing. High Q-factors of surface plasmon modes in VNTs translate into high spectral resolution of sensors. In particular, Fig. 12.16 demonstrates that novel bio(chemical) sensors based on the plasmonically-integrated tornadoes may offer at least an order of magnitude improvement in the figure-of-merit values as compared to current designs. In Fig. 12.16, the performance of the VNT in Fig. 12.15 as a refractive index sensor is evaluated by using the standard figure of merit: $FoM = sensitivity/linewidth$ [106, 107]. Sensor sensitivity is defined as the resonance shift resulting from the ambient refractive index change: $S[nm/RIU] = \Delta\lambda/\Delta n$. The 'linewidth' of an asymmetrical resonant feature is measured as the wavelength difference between the resonant peak and the closest neighboring dip [106]. The VNT scattering spectrum shows pronounced redshifts with the refractive index increase (Fig. 12.16a). The linear regression slope for the resonance shift plotted in Fig. 12.16b yields the sensitivity value 518 nm/RIU, on par with those reported for other plasmonic nanosensors. However, owing to the narrow linewidths of the resonant features, the resulting FoM value of 85 is almost an order of magnitude larger than those predicted and measured in single particles (from ~1 to ~5.5) [107, 108] and nanoclusters (up to ~11) [106, 108]. Using low-loss nanoshells [109] and optimizing VNT configurations could enable even higher FoM values. Furthermore, the strong localized fields generated in VNTs provide optical trapping potentials strong enough to capture low-index particles and biological macromolecules. This offers novel solutions for background-free sensing of optically-trapped nano-objects, potentially reaching single-molecule sensitivity. It should also be noted that the high sensitivity of the VNTs optical spectra to the substrate stretching (Fig. 12.11c,d) makes them ideal candidates for nanoscale stress and strain sensors.

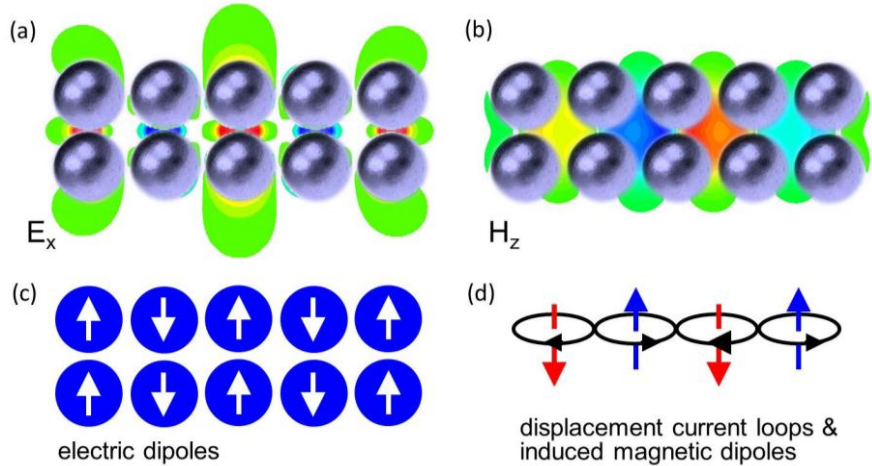

Fig. 12.17. Magnetic effects in vortex nanogears transmissions. Electric (a) and magnetic (b) field distributions in the plane of the VNT shown in Fig. 12.10 at the frequency of its collective mode resonance. Schematics of the induced electric dipoles (c), oscillating displacement currents and magnetic dipole moments (d).

The proposed approach also offers the promise of a broad impact on nanoplasmonic-based renewable energy applications as it helps to eliminate the mismatch between the electronic and photonic length scales in thin-film photovoltaic devices. Most importantly, the expected increase of the efficiency and spectral band-width of light absorption can be achieved with the simultaneous reduction of the dissipative losses in metals. Light harvesting platforms based on folding optical powerflow into nanoscale vortices may help to minimize the thickness of semiconductor needed to absorb light completely and to amplify the signal via plasmonic enhancement mechanism. They will also be compatible for integration with either silicon electronics or flexible substrates such as those based on organic and polymer materials.

Finally, excitation of collective coupled-vortex resonances in VNTs is accompanied by the generation of circulating displacement currents (Fig. 12.17). These currents, according to Maxwell's equations (12.1), induce magnetic moments (Fig. 12.17d) in the same way as circulating conducting currents in wire loops generate time-varying magnetic fields [33, 109, 110]. This effect can be clearly seen in Fig. 12.17, where the electric and magnetic field distributions are shown in the VNT plane. As schematically shown in Fig. 12.17d, antiphase magnetic moments are generated in the adjacent loops of displacement current, and their mutual coupling along the chain gives rise to an artificial "antiferromagnetic" behavior. We can now conclude that, according to Eqs. (12.4,12.5) the observed reduction of dissipative losses in VNTs in comparison to conventional plasmonic designs is achieved because (i) the electromagnetic energy is re-circulated outside of metal volume and thus does not get converted in to the kinetic energy of electrons, and (ii) noticeable portion of the energy is stored in the magnetic rather than electric field. Excitation of magnetic plasmons has also been observed in fused heptamer plasmonic molecules shown in Fig. 12.2c [44], however, in that case the energy is re-circulated through the metal volume, which increases dissipative losses. We have also recently demonstrated magnetic response in hybrid metal-dielectric vortex-pinning structures [111].

## 12.7. Conclusions and outlook

A new hydrodynamics-inspired approach to plasmonic nanocircuit design is discussed. It helps to alleviate major setbacks to their practical applications, such as high radiative and/or dissipative losses of noble-metal nanostructures in the visible frequency range. In the frame of the new approach, basic building blocks of plasmonic nanocircuits are designed to feature phase singularities in the optical near fields. This results in the creation of areas of circulating electromagnetic energy flow – optical vortices – at pre-defined positions. As the examples discussed in this chapter demonstrate, the individual vortex-pinning building blocks may not exhibit any interesting electromagnetic behavior (such as high field intensity or narrow frequency linewidths). As such, they would have been discarded in the frame of the traditional design approaches. However, tailored coupling of individual nanoscale vortices into vortex nanogear transmissions may yield dramatic optical effects never observed before. The data presented in Fig. 12.18 summarize the advantages offered by the proposed rational design strategy. It offers a com-parison between the optical performance of several popular designs of plasmonic components, including a nanodimer, a heptamer plasmonic molecule, and a periodic nanoparticle chain with the VNT structures shown in Figs. 12.10 and 12.14. For consistency, all the nanostructures in Fig. 12.18 are composed of identical 100-nm Ag nanoparticles and have the same minimum inter-particle gaps of 3nm. Fig. 12.18 shows that rationally-designed VNTs offer the way to achieve previously unattainable high-Q plasmonic modes within sub-wavelength footprints.

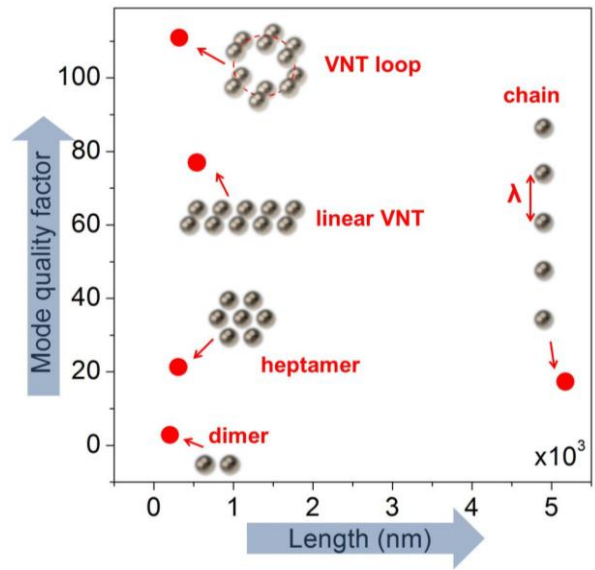

Fig. 12.18. Design roadmap for achieving high-Q modes in sub-wavelength plasmonic nanostructures composed of 100nm silver nanoparticles.

The advantages of the new vortex-pinning plasmonic platforms for sensing and spectroscopy having been demonstrated, other applications can be explored in the future. These include renewable energy generation (i.e., photovoltaic and photocatalytic platforms), metamaterials design, and reconfigurable nonlinear plasmonic nanocircuits [112, 113]. Another future exciting application for plasmonically-integrated nanoscale tornadoes is in optical trapping of small particles (viruses, bacteria, etc) and guiding them through on-chip nanoscale 'conveyor belts' [114]. The insights into the underlying physical mechanisms and analogies with the well-studied hydrodynamic effects are expected to help in developing additional design rules for complex plasmonic nanocircuitries. The hydrodynamics analogy also paves the road to using well-developed methods of computational fluid dynamics to facilitate simulation and optimization of plasmonic VNTs.

## Acknowledgements

I would like to thank Dr. Anton Desyatnikov from Australian National University and my colleagues at Boston University and MIT for useful discussions.